\documentclass[12pt]{iopart} 
\pdfoutput=1
\usepackage{graphicx} 
\usepackage[caption=false]{subfig}
\usepackage{color}
\usepackage{float}

\begin{document}

\title[]{Assessing Spatial Information in Physical Environments}

\author{Vinicius M. Netto$^{\ominus}$, Edgardo Brigatti$^{\star}$, Caio Cacholas$^{\ddag}$ and Vinicius Gomes Aleixo$^{\ddag}$}

\address{$^{\ominus}$ Department of Urbanism, Universidade Federal Fluminense (UFF), Rua Passo da Patria 156, Niteroi, Rio de Janeiro state, Brazil} 
\address{$^{\star}$ Instituto de Física, Universidade Federal do Rio de Janeiro, Av. Athos da Silveira Ramos, 149, Cidade Universitaria, 21941-972, Rio de Janeiro, RJ, Brazil}
\address{$^{\ddag}$ Programme of Graduate Studies, Universidade Federal Fluminense (UFF), Rua Passo da Patria 156, Niteroi, Rio de Janeiro state, Brazil}

\maketitle

\begin{abstract}
Many approaches have dealt with the hypothesis that the environment contain information, mostly focusing on how humans decode information from the environment in visual perception, navigation, and spatial decision-making. A question yet to be fully explored is how the built environment could encode forms of information in its own physical structures. 
This paper explores a new measure of spatial information, and applies it to twenty cities from different spatial cultures and regions of the world. Findings suggest that this methodology is able to 
identify similarities between cities, generating a classification scheme that 
opens up new questions about what we call 'cultural hypothesis': the idea that spatial configurations find consistent differences between cultures and regions.
\end{abstract}

\section{Introduction}

One of the most interesting question in spatial information is \textit{how the built environment could carry forms of information encoded in its own physical structures}. Of course the idea that cities can actually encode information is well known in urban theory. It is at the heart of Kevin Lynch's \cite{lynch1960image} pioneering work in 1960 on spatial elements guiding navigation in the environment, along with memory and representation, even though he did not quite use the term 'information'. In the 1980s, Amos Rapoport \cite{rapoport1982meaning} (page 19) explicitly asserted, "physical elements of the environment do encode information that people decode". 
More recently, Haken and Portugali \cite{haken2003face,haken2014information} have seen information latent in street layouts and built form. 
The possibility that information could be encoded in physical structures is indeed very interesting. 
If physical spaces materialise information, 
we could encode information in the built environment and decode it while living in it. Information materialised in physical space could be contextual resources useful to guide our actions. 
Research in spatial information has certainly dealt with this possibility focusing on how we decode information from the environment
\cite{tanner1954decision,garlandini2009evaluating}. 

A question yet to be fully explored is \textit{how the built environment could encode forms of information in its own physical structures} in the first place, and \textit{how such amounts of information could be measured empirically}. Approaches to information have adopted different measures
to try to extract task relevant information \cite{leibovici2009defining, woodruff1998constant,rosenholtz2007measuring}. In this paper, 
we further develop and apply a new approach \cite{netto2018cities} based on Shannon entropy to assess urban configurations as environmental information. 
With this aim in mind,
we characterise the spatial information encoded in two-dimensional configurations of buildings, seen as the most elementary unit of the urban fabric. 
By analysing the cellular arrangement of buildings, we are able to assess the structures of urban blocks and streets. 
Information will be quantified measuring Shannon entropy \cite{shannon1948_1,shannon1948_2}, operationally estimated by looking at the sequence of bits 1 and 0 representing built form cells and open space cells within sections of cities.
Theoretically, this corresponds to analysing a 2D symbolic sequences of 1 and 0. In this context, information finds a very precise meaning:
the entropy of the sequence, a measure of the surprise a source that produces the sequence causes in the observer \cite{entropy90}. 
In fact, physical arrangements  characterised by higher levels of randomness, uncertainty or unpredictability  are associated with high entropy. 
In contrast, the presence of regularities and patterns in urban structures will correspond to lower entropy, which means a higher predictability.
Hypothetically, cities with ordered structures would help agents understand their environment, allowing them to make predictions about areas beyond their fields of visibility. Agents can make inferences, memorise layouts and navigate more easily from one place to another—say, grasping the pattern of blocks and intersections from local streets and inferring that some blocks away they will still have the same pattern (see \cite{lynch1960image, hillier1996space, hillier2012genetic, boeing2018urban,louf14} on legibility and intelligibility in
 urban structures; \cite{portugali2011complexity, tanner1954decision, garlandini2009evaluating} on pattern recognition and spatial decision making).

Indeed, the layout of the environment encodes more information than two-dimensional configurations can express. However, 
we opted for an analytic approach able to sufficiently describe differences in information potentially encoded in urban built form -- hence our reduction of urban form to cellular aggregations.
We apply this approach to a number of empirical cases, namely 20 cities from three different regions of the world. We expect our measure 
to grasp 'spatial signatures' of such cities, i.e. a measure able to point out differences and quantify similarities
inherent to their spatial configurations. 
Thus we introduce a general concept of similarity between pairs of cities based on their informational content.
This concept will allow to order our pool of cities and define a classification scheme  which can help in verifying something that captures the imagination of many scholars: the possibility of finding consistent similarities in the configuration of cities from a same 'spatial culture' or world region, along with consistent differences between cities from different cultures -- what we call \textit{'the cultural hypothesis'}. 

\section{Materials and methods}

Our database includes 20 cities from the South of Europe, North America 
and Latin America. We selected cities based on (i) their importance for the region and country; (ii) a size compatible with our methodological requirements; (iii) availability of data. 
The first item has brought us to some well-known cities as emblematic cases. 

The second aspect involved the selection of areas for the application of our measure. 
Selection of cities and sections was based on the identification of homogeneous areas, i.e. with spatial continuity of its urban fabric able to satisfy occupation rates close to 50\%, 
 which means avoiding large empty areas or rarefied patterns of urbanisation.  
These restrictions follow two critical considerations. The first is that our method is well fitted for estimating entropy for continuous urban areas. Second, and most importantly, it is interesting to decouple the analysis of city structures between small-scale, detailed and denser urban areas, and large-scale regional and peripheral urban areas.  
In fact, the two scales are different, and for this reason, they can be described using different methodologies. The first one is marked by specific features such as blocks and buildings, and the stratification of human interventions uniquely characterises resulting shapes and structures.  
The second scale is distinguished by sparse occupation, frequently with a scale-free character, where physical phenomena and constraints that geographical formations and barriers might play very relevant roles. 
In this work we will focus only on small scale, continuous urban areas.


The third criterion involved availability of spatial information on the configuration of cities. Many cities, particularly in Latin America, 
have incomplete information regarding buildings and their precise location, position and geometry (for instance, major cities like Lima and Bogota). 

We extracted sections of cities from public map repository Google Maps API. We tested trade-offs between resolution and availability of data for distinct scales. We chose geographic areas of $9,000,000$ m$^2$, which were considered sufficient for representing the general spatial characteristics of dense urban areas regarding the configuration of buildings, urban blocks and street networks. 
Background picture bases of the selected cities were then prepared and exported in high resolution, filtering layers and converting entities representing buildings into solid raster cells.  Images underwent a re-sizing process for $1000^2$ cells and were converted to a monochrome system and then into a matrix of size $1000\times 1000$ cells with binary numerical values.\\ 

\begin{figure}[b!]
    \centering
    \includegraphics[width=.65\textwidth]{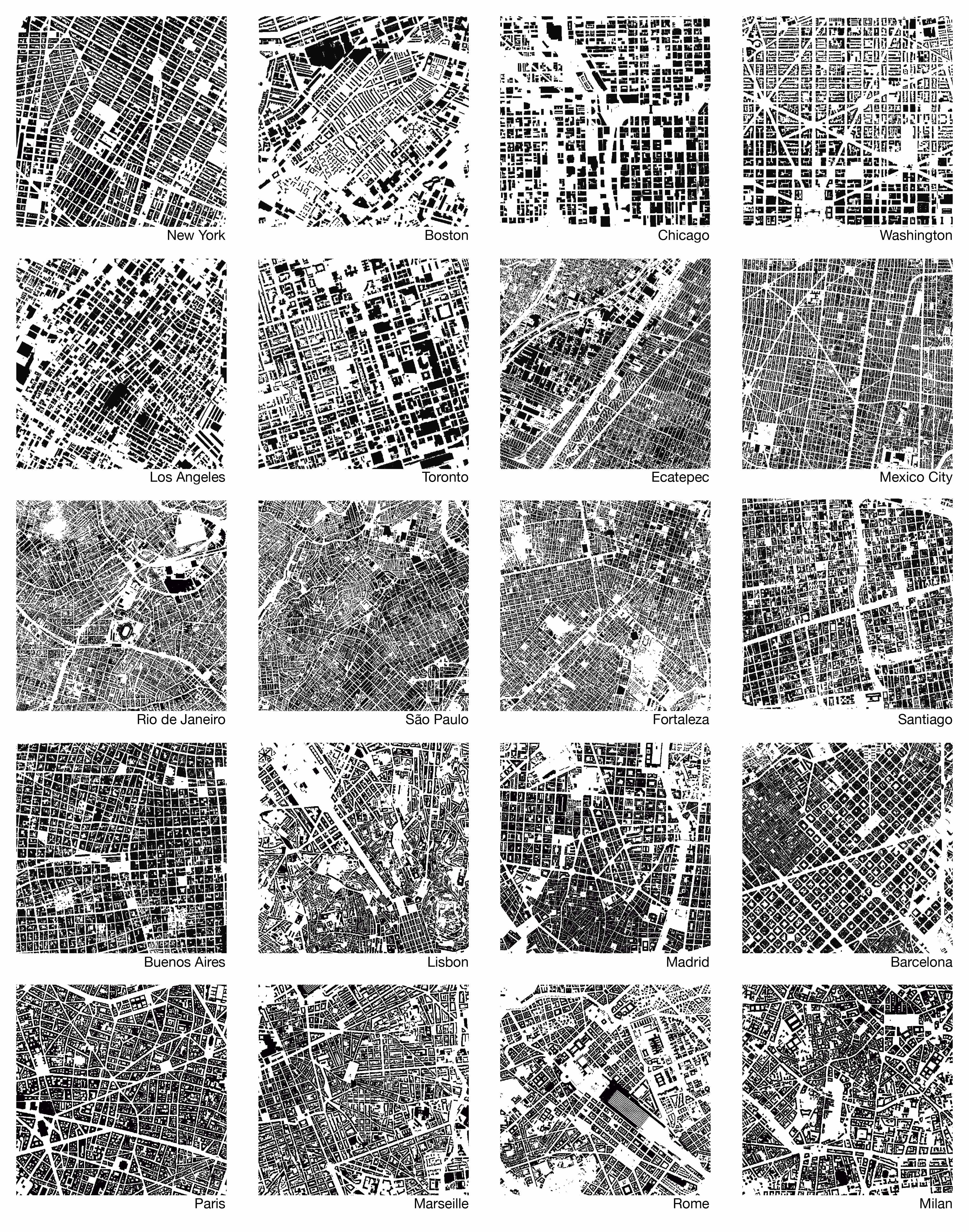}
    \caption{\small Spatial distributions in real cities ($9,000,000$ m$^2$ windows, 1000×1000 cells), extracted from Google Maps. These sections of emblematic cities are used to compute Shannon entropy. 
    }
    \label{fig:figure6}
\end{figure}

We propose to assess Shannon entropy 
in the 
cellular arrangements displayed in figure \ref{fig:figure6}. Our approach uses a method generally applied for estimating the entropy of sequences of symbols encoded in one-dimensional strings \cite{schurmann1996entropy}. 
For these data sets, 
the method consists of defining the block entropy of order $n$ through

\begin{equation}
H_n=-\sum_k  p_n(k) \log_{2}[p_n(k)],
\label{1entropy}
\end{equation}
{where blocks are string segments of size $n$, and the  sum runs over all the $k$ possible $n$-blocks.
Equation~(\ref{1entropy}) corresponds to the Shannon entropy of the probability distribution $p_n(k)$.}
The Shannon entropy 
of the considered system \cite{schurmann1996entropy, lesne2009entropy}, which we indicate with $h$, is
obtained from the following limit:

\begin{equation}
h=\lim_{n \to \infty} H_n/n,
\label{hentropy}
\end{equation}

which measures the average amount of randomness per symbol that persists after all correlations and constraints are taken into account.
It was proven that the above limit exists for all spatial-translation invariant systems.
More details about this approach can be found in \cite{schurmann1996entropy,lesne2009entropy}.

\begin{figure}[ht]
    \centering
    \includegraphics[width=0.70\textwidth]{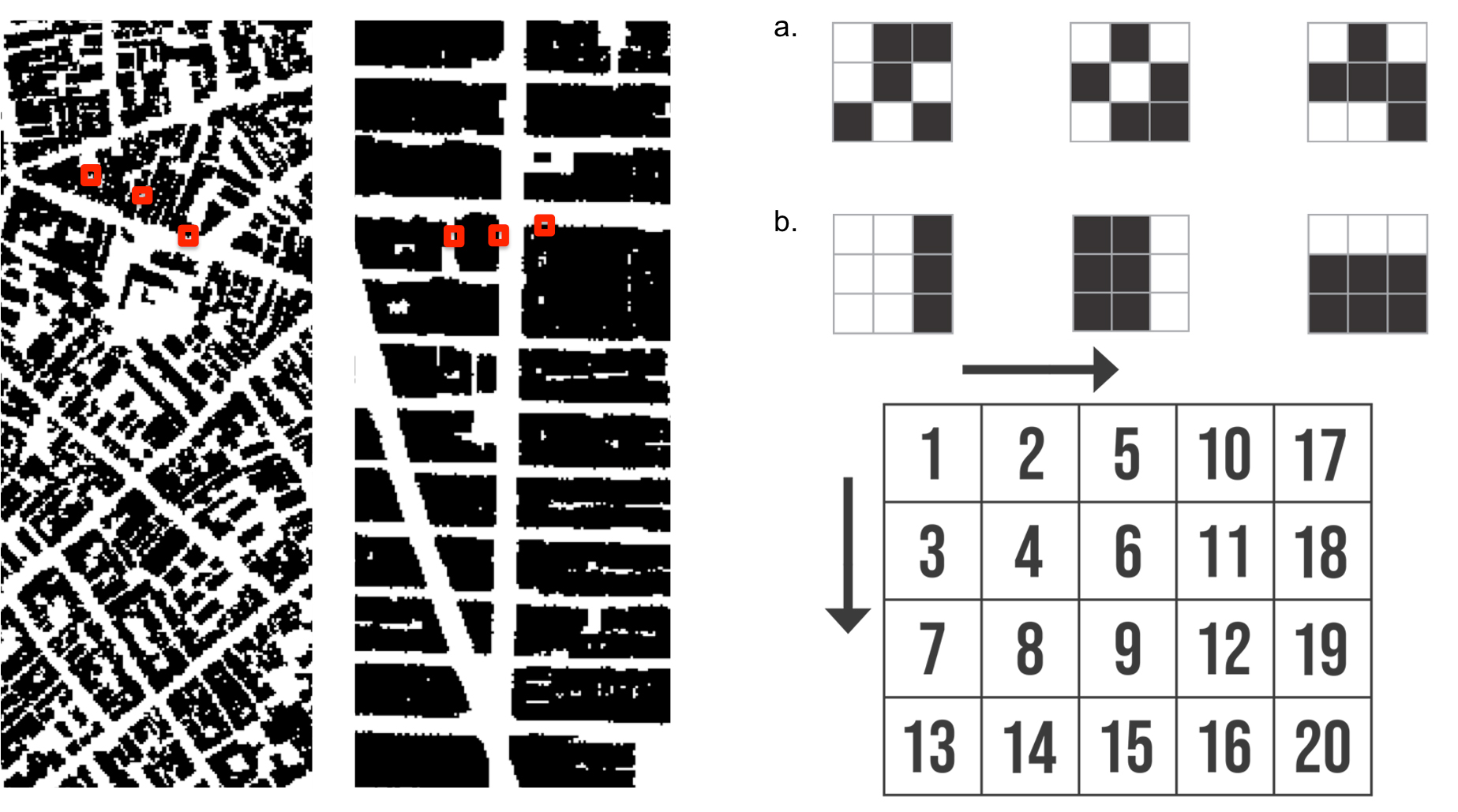}
    \caption{
    Examples of blocks with nine cells are shown in red for selected areas in Rio and Manhattan, NY (left), and are amplified on the right. 
    Blocks are constructed following the determined 1-D path represented on the bottom, right. Numbers indicate the order in which the cells are added to the block. The first block of size 1 corresponds to cell 1 and neighbouring cells are added in the corresponding order. 
    }
    \label{fig:blocks+20cells1.jpg}
\end{figure}

This method can be applied to our problem once we have defined the blocks for a two-dimensional matrix \cite{feldman2003structural}. In this two-dimensional context, the most intuitive idea is to consider a block of size $n$ as a square which contains $n$ cells. 
To obtain the sequence of $H_{n}$ also for $n$ values that do not correspond to squares, we considered blocks that interpolate perfect squares, as described in figure \ref{fig:blocks+20cells1.jpg}.
Note that there is no unique natural way to scan a 2D matrix. 
We tested our approach for
different reasonable forms of constructing the blocks, and using different paths does not seem to strongly influence the estimation of $H_n$.

Equation \ref{hentropy} gives precisely the entropy for a theoretical infinite set of data.
In real situations, where the data set is finite, our method estimates the probabilities of distinct arrangements of cells within blocks up to a certain size $n$, counting their frequencies, and then estimates the limit.
Note that when working with two symbols, the estimation of $H_n$ becomes not reasonable when $2^{n}\approx N$, where $N$ is the number of elements in our data set \cite{lesne2009entropy}. 
Thus, in our case, this condition is verified for $n\approx 20$.

\section{Results and discussion}

\begin{figure}[h!]
\begin{center}
\includegraphics[width=0.90\textwidth]{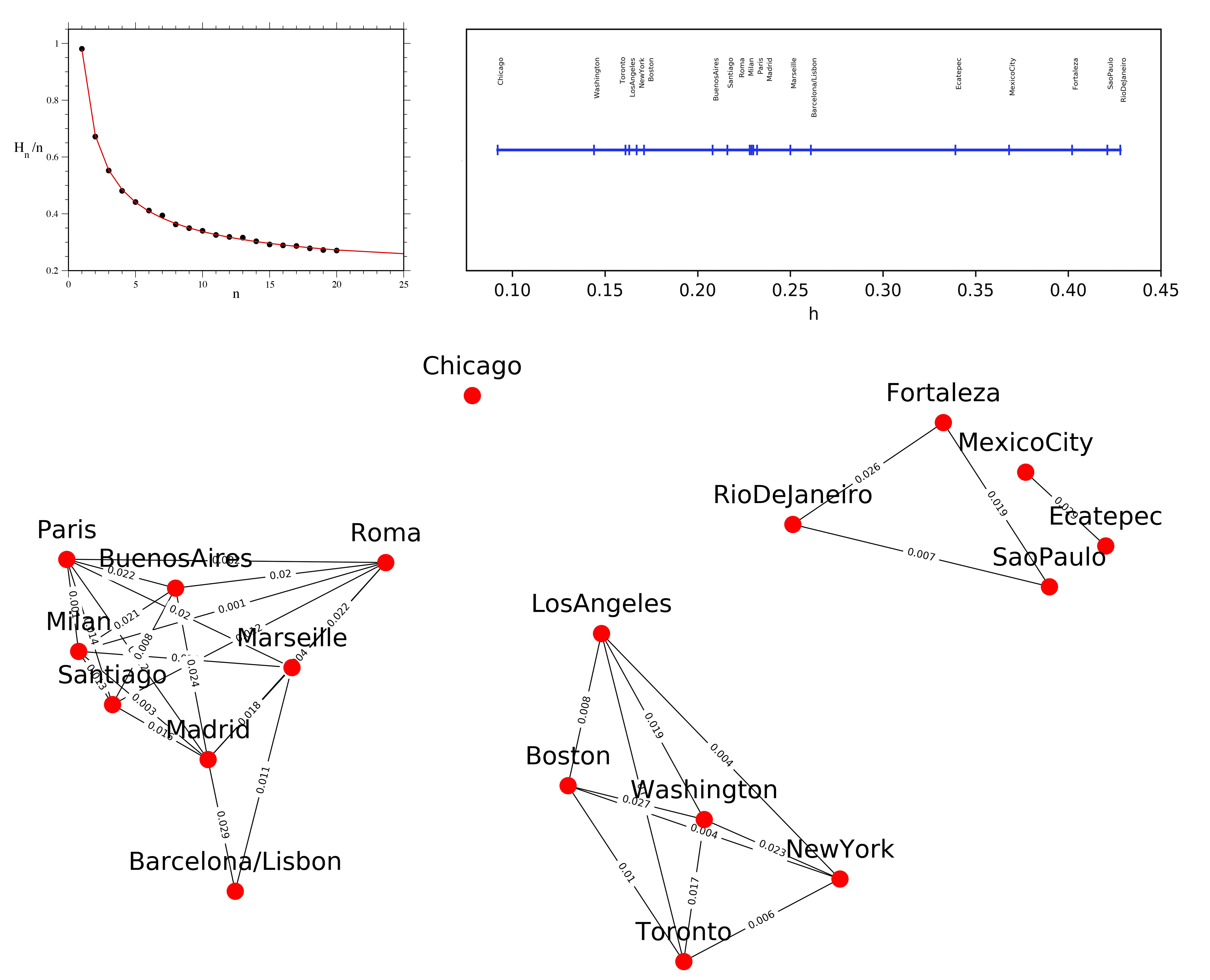}
\end{center}
\caption{\small \textbf{top left}: An example of the estimated values of $H_n/n$ for the city of Los Angeles. The continuous line represents the best fitting of our data using the function: $a+b/n^c$. The fitted values of $a$ give an extrapolation of the Shannon Entropy $h$ of the data set. 
All the analyzed cities present a very similar behaviour.
\textbf{top right:} Estimated values of $h$ for cities under analysis.\\
\textbf{bottom:} Proximity network of cities based on the value of $h$.
The edge lengths are not proportional to the levels of proximity between entropy values. 
Distances are labelled over edges.}
\label{fig:tres}
\end{figure}

We found empirically that, for all examined cases, the following ansatz provides an excellent fit:
$a+b/n^c$. 
Even if we observed that the convergence is relatively slow, the fitted value of $a$ gives a reasonable extrapolation of the Shannon Entropy $h$.
As an example, we show the results for the estimation of $H_n$/$n$ and the corresponding entropy $h$ for the city of Los Angeles in figure \ref{fig:tres} (top left).
Results for the estimation of the entropy $h$ for the selected areas of the twenty sampled cities are shown in Table \ref{table0}.

\begin{table}[h!]
    \centering
{\scriptsize
    \begin{tabular}{|c|c|c|c|} 
\hline
\hline
 Chicago   & 0.092 &     Paris      & 0.230 \\
\hline
 Washington & 0.144 & Madrid    & 0.232 \\
\hline
Toronto        & 0.161 & Marseille   &  0.250 \\
\hline
LosAngeles  & 0.163 & Barcelona & 0.261\\
\hline
NewYork       & 0.167 & Lisbon & 0.261 \\
\hline
Boston          & 0.171 & Ecatepec  & 0.339 \\
\hline
Buenos Aires & 0.208 & Mexico City &  0.368 \\
\hline
Santiago       & 0.216 & Fortaleza    & 0.402 \\
\hline
Roma            & 0.228 & S\~ao Paulo &  0.421 \\
\hline
Milan         & 0.229 & Rio de Janeiro   & 0.428 \\
\hline    
\hline
\end{tabular}
}
\caption{\small The estimated  entropy $h$ for selected areas of the 
sampled cities.
The errors relative to the fitting procedure are of the order of $0.01$.
}
\label{table0}
\end{table} 

We proceeded to analyse these values in search of similarities and differences between the entropy signatures of the sampled cities. For clarity, we initially plotted the entropy values along a straight line  (figure \ref{fig:tres}, top right).
Aiming to use these results for developing a classification scheme for our dataset, 
we performed a proximity network analysis based on the entropy values, to identify the possible formation of clusters of cities sharing similar values. Once we  obtained the entropy $h$ for all considered cities, we can
quantify the levels of similarity 
defining a distance  between cities $i$ and $j$ based on the values of $h$:
$d_{ij}=|h_i-h_j|$.
We created a matrix of distances for the twenty cities and then defined a network where cities are nodes, and edges (links between nodes $i$ and $j$) are present only if the value of $d_{ij}$ is smaller than a fixed threshold value of $0.03$, which roughly corresponds to the 99\% C.I. of the extrapolated values of $h$.
The proximity network of cities 
can be seen in figure \ref{fig:tres}.

We can clearly distinguish between five different, disconnected clusters.
In the lower side of the cellular entropy spectrum, Chicago is the most ordered structure. It is the case to an extent that it appears on a class of its own, an isolated node with an entropy value $h$ below 0.100.
We have a strong cluster formed by other cities selected in North America (US/Canada), with $h$ from 0.144 (Washington) to 0.171 (Boston).
South American cities Buenos Aires and Santiago follow in a different cluster (entropy values above 0.200), along with European cities.
Since other cities in Latin America have two clusters close in entropy levels
(one with cities in Brazil and one with the cities in Mexico), this is an apparently surprising result -- one that runs counter the cultural hypothesis of similarity in spatial signatures for cities within a same culture or region. This suggests that a single, general regionally-based explanation cannot respond to all cases. In fact, deeper historical contingencies including specific cultural factors might be forces at play. 
For instance, cities founded by Spanish colonizers in Latin America were often created as an orthogonal grid pattern, which were the case for Buenos Aires and Santiago. As these cities expanded, patchwork patterns were added around the central core's regular structure, adding entropy to the mix. Nevertheless, order in those spatialities is still felt today.
Close to them, Southern European cities appear as a middle range cluster, with $h$ from 0.228 (Rome) to Lisbon (0.261). Of course these cities have fully functional systems of urban blocks and street networks even though they are arranged as deformed grids, probably due to centuries of urbanization since the Middle Ages and ancient foundations. 
Finally, we have the cluster on the higher side of the cellular entropy spectrum, with South American cities in Brazil (Rio de Janeiro, Sao Paulo and Fortaleza). 
This finding is consistent with Medeiro's \cite{Medeiros2013} analysis of Brazilian cities as in average the most topologically fragmented among 164 cities in the world, and with Boeing's \cite{boeing2018urban} finding about S\~ao Paulo having one of the three lower grid-order values in a sample with a hundred cities (which did not include Rio).

To sum up, in this paper, we attempted to understand how spatial information is encoded in built environments, looking 
into their physical configurations as cellular aggregations. 
We proposed a measure of physical information, part of a larger three-layered model of environmental information-interaction \cite{netto2017social}, and we applied this measure to a sample of twenty cities from different world regions. We found empirical signs that our entropy measure 
is powerful enough to capture consistent 'spatial signatures' of different cities.

Although our sample is limited, 
our results are in general consistent with previous findings based  on different spatial measures. Chicago and American/Canadian cities appear prominently as the most ordered in their cellular configurations, as they are once assessed as metrical, topological or fractal systems \cite{Medeiros2013, rashid2017geometry}, or in terms of grid order \cite{boeing2018urban}. But there are interesting specificities identified by our approach.
South American cities Buenos Aires and Santiago appear in a cluster with European cities. Since other cities in South America have a cluster of their own, this is an apparently surprising result -- one that runs counter the cultural hypothesis. 

Our method and findings about physical information signatures of different cities are equally suggestive considering the hypothesis of cultural and regional similarities that lies in the urbanistic imagination. On the one hand, \textit{a general regionally-based theory} of regional morphogenesis of structurally similar cities seems unable to respond to every individual case. 
Deeper historical contingencies including specific cultural factors might actively shape cities. 
On the other hand, our results do not allow us to simply refute the cultural hypothesis. In general, the hypothesis stands, but it does require a more complex consideration: \textit{culture matters, but cannot be reduced to regional borders}.\\

\appendix

\bibliographystyle{plainurl}
\bibliography{overleaf2019}

\appendix

\end{document}